\documentclass[11pt,a4paper]{article}

\usepackage{epsfig,amsmath,amssymb,cite}
\def\lp{\left(}
\def\rp{\right)}

\def\be{\begin{equation}}
\def\ee{\end{equation}}
\begin{document}

\title{Asymptotically anti-de Sitter cylindrical thin-shell wormholes} 
\author{Ernesto F. Eiroa$^{1,2,}$\thanks{e-mail: eiroa@iafe.uba.ar}, 
Claudio Simeone$^{2,3,}$\thanks{e-mail: csimeone@df.uba.ar}\\
{\small $^1$ Instituto de Astronom\'{\i}a y F\'{\i}sica del Espacio (IAFE, CONICET-UBA),} \\ {\small Casilla de Correo 67, Sucursal 28, 1428, Buenos Aires, Argentina}\\
{\small $^2$ Departamento de F\'{\i}sica, Facultad de Ciencias Exactas y 
Naturales,  Universidad de Buenos Aires,} \\ 
{\small Ciudad Universitaria Pabell\'on I, 1428, 
Buenos Aires, Argentina}\\ 
{\small $^3$ IFIBA--CONICET, Ciudad Universitaria Pabell\'on I, 1428, 
Buenos Aires, Argentina}} 
\maketitle

\begin{abstract}

In this article we investigate cylindrical thin-shell wormholes which are asymptotically anti-de Sitter. We analyze their stability under perturbations preserving the symmetry by using two different methods. We compare the results with those corresponding to the wormholes constructed from the Levi-Civita spacetime. We find that the configurations always require the presence of exotic matter at the throat, and, in the case of the linearized stability analysis, they can be stable for suitable values of the parameters.\\

\noindent 
PACS number(s): 04.20.-q, 04.20.Gz, 04.20.Jb 

\end{abstract}

\section{Introduction}

Traversable Lorentzian wormholes were first considered by Morris and Thorne \cite{mo}; because of the interesting unusual properties of these topologically non trivial geometries, in the years after this leading work they have received considerable attention \cite{visser}. A major objection against their actual existence, however, is the necessity of ``exotic'' matter (not fulfilling the energy conditions) supporting the wormholes; within the framework of general relativity this difficulty cannot be avoided in the case of compact wormhole geometries. Therefore, wormholes of the thin-shell class \cite{povi1,povi2,visser}, that is, those which are mathematically constructed by cutting and pasting two geometries, present a particular interest because in such configurations the exotic matter can be restricted to a layer placed at the throat (see Refs. \cite{whsphe1,whsphe2} and references therein). 

In the last three decades, cosmic strings \cite{cs} have been the object of a thorough study, because of their possible relevance in the process of structure formation in the early Universe \cite{structure}, and also because, in principle, they could be detected by gravitational lensing effects \cite{lens}. Besides,  open or closed fundamental strings are the center of present day programs in the search for a unified theory of the fundamental interactions. Then the interest in the gravitational effects of both fundamental and cosmic strings and, in general, in axially symmetric solutions of the equations of gravitation, has been recently  renewed   (for instance, see Ref. \cite{strings}). Consequently, cylindrically symmetric wormholes have been studied in the last few years; see Refs. \cite{cle,aros,ku,brle,ei1,ei2,mc,eisi10,cms,sharif,ri13,melvin}.  
In particular, a remarkable aspect was noted in Ref. \cite{brle}:  cylindrically symmetric wormholes admit two definitions of the throat, and one of them --that the geodesics {\it restricted to a plane normal to the symmetry axis} open up-- can be compatible with the energy conditions even within the framework of general relativity. It has been shown that such cylindrical wormholes satisfying the weak energy condition cannot have a flat or local cosmic string asymptotic behavior at both sides of the throat. Wormholes associated with an azimuthal magnetic field and supported by non exotic matter were found; they are neither symmetric with respect to the throat nor flat or local string like at infinity (see Ref. \cite{brle} and references therein). 
The possibility of cylindrical thin-shell wormholes supported by a layer of positive energy density placed at the throat was proved in Ref. \cite{eisi10} within the framework of Einstein's gravity. Besides, the linearized stability of generic thin-shell wormholes was studied in Ref. \cite{mha}.

Cylindrical spacetime solutions with non vanishing cosmological constant $\Lambda$ were found in Refs. \cite{linet,tian}; they are known as the Linet--Tian metric or the Levi-Civita  metric with a cosmological constant, and are associated to the exterior of a cosmic string.  Several interesting features of these geometries were analyzed in Refs. \cite{zobi,swps,bsms,gp}. A case of particular interest is the one corresponding to a negative cosmological constant, which is asymptotically anti-de Sitter. In recent years, anti-de Sitter spacetimes have been of interest  within the context of supergravity and string theories; in particular, they play an important role in M-theory. The AdS/CFT correspondence conjecture provides a link between gravity and conformal field theory. The Linet--Tian metric with  $\Lambda < 0$ has the form \cite{linet,tian,zobi}:
\begin{equation}
ds^2=-Q(r)^{2/3} P(r)^{-2 \zeta (k)/ 3 \alpha (k) }dt^2+ dr^2+ Q(r)^{2/3} \left[ \xi ^{-2} P(r)^{2\epsilon (k)/3 \alpha (k) }d\varphi^2 
+ P(r)^{2 \gamma (k)/3 \alpha (k)}dz^2 \right],
\label{ds2}
\end{equation}
where
\begin{equation*}
P(r)= \frac{2}{ \sqrt{3|\Lambda |}} \tanh \left( \frac{\sqrt{3|\Lambda |}r}{2}\right),
\end{equation*}
\begin{equation*}
Q(r)=\frac{1}{\sqrt{3|\Lambda |}} \sinh \left( \sqrt{3|\Lambda |} r\right),
\end{equation*}
$\alpha (k) =4 k^2-2 k+1$, $\zeta (k)=4 k^2-8 k+1$, $\epsilon (k)= -4k^2-4k+2$, and $\gamma (k)= 8 k^2-4 k-1$.  The constant $\xi$ is related to the angle defect, and the  parameter $k$ is related, but not equal, to the linear mass density of the source. When $\Lambda\to 0$, we have that $P(r)=Q(r)=r$ and one recovers the Levi-Civita spacetime \cite{lc}. For the parameter $k$, the physically relevant range  \cite{zobi,bsms}  is $0 \le k \le 1/2$. The solution (\ref{ds2}) is singular in the axis $r=0$, except for the cases $k=0$ and $k=1/2$. When $k=0$ the geometry corresponds to the so-called non uniform anti-de Sitter universe, which is one of the two known vacuum solutions of the Einstein equations with a negative cosmological constant \cite{bon}. If $k=1/2$, it has been shown \cite{swps,zobi} that with a suitable coordinate change the spacetime corresponds to a black string.

Cylindrical thin-shell wormholes were constructed from the Linet--Tian metric with a positive cosmological constant  \cite{ri13,mha}, for which the stability was recently studied.  In the present article, we consider the mathematical construction and characterization of cylindrically symmetric thin-shell wormholes with a negative cosmological constant; in particular,  a detailed analysis of their stability under perturbations preserving the symmetry is performed. Two different approaches for the stability analysis are adopted: in the first one, we assume that the equations of state for the static configuration are preserved along the evolution of the perturbation, and in the second one, we assume a linearized equation of state relating each pressure with the surface energy density. We compare the results with those corresponding to the thin-shell wormholes obtained from the Levi-Civita metric.  We adopt units such that  $G=c=1$.

\section{Wormhole construction}\label{cylind}

The mathematical construction of generic thin-shell wormholes with cylindrical symmetry (see Ref. \cite{eisi10}) starts from a metric which, in coordinates $X^{\alpha}=(t,r,\varphi ,z)$, has the form 
\begin{equation}
ds^2 = -A(r)dt^2 +B(r)dr^2 +C(r)d\varphi ^2+D(r)dz^2;
\label{e1}
\end{equation}
here $A$, $B$, $C$, and $D$ are positive functions of the radial coordinate only. Two copies $ \mathcal{M}^{\pm} = \{ X / r \geq a \}$ of the region with $r \geq a$ are taken, and then joined at the hypersurface
$ \Sigma \equiv \Sigma^{\pm} = \{ X / r - a = 0 \}$. The resulting new manifold $\mathcal{M}=\mathcal{M}^{+} \cup \mathcal{M}^{-}$ is geodesically complete. If a flare-out condition is satisfied, that is, if the geometry opens up at the radius $a$, the construction determines a cylindrically symmetric thin-shell wormhole with two infinite regions connected by a throat at $\Sigma $.  We apply the standard Darmois-Israel formalism \cite{dar,israel, mus}.
At the throat we adopt coordinates $\xi ^i=(\tau , \varphi,z )$, with $\tau $ the proper time on the shell. The radius of the throat is, in general, a function of $\tau $, that is,  $a = a ( \tau )$; in this case, the shell is defined by $\Sigma : \mathcal{H} ( r, \tau ) = r - a ( \tau ) = 0$. The energy density and pressures of matter on the shell are related with the extrinsic curvature $K_{ij}$ at the two sides of it.  The relation is given by the Lanczos equations 
\begin{equation}
-[K_{ij}]+[K]g_{ij}=8\pi S_{ij},
\label{e7}
\end{equation}
where $[K_{ij}]\equiv {K_{ij}}^+ - {K_{ij}}^-$, 
$[K]=g^{ij}[K_{ij}]$ is the 
trace of the extrinsic curvature tensor and $S_{ij}$ is the surface stress-energy tensor. If we define the unit normals $n_{\gamma}^{\pm}$ (so that $n^{\gamma} n_{\gamma} = 1$) to the surface 
$\Sigma$ as
\begin{equation}
n_{\gamma}^{\pm} = \pm \left| g^{\alpha \beta} \frac{\partial
\mathcal{H}}{\partial X^{\alpha}} \frac{\partial \mathcal{H}}{\partial
X^{\beta}} \right|^{- 1 / 2} \frac{\partial \mathcal{H}}{\partial
X^{\gamma}}, \label{e3}
\end{equation}
then the extrinsic curvature at each side of the shell is given by
\begin{equation}
K_{ij}^{\pm} = - n_{\gamma}^{\pm} \left. \left( \frac{\partial^2
X^{\gamma}}{\partial \xi^i \partial \xi^j} +\Gamma_{\alpha\beta}^{\gamma}
\frac{\partial X^{\alpha}}{\partial \xi^i} \frac{\partial
X^{\beta}}{\partial \xi^j} \right) \right|_{\Sigma}. \label{e2}
\end{equation}
Working in the orthonormal basis $\{ e_{\hat{\tau}} = \sqrt{1/A(r)}e_{t}$, $e_{\hat{\varphi}} 
= \sqrt{1/C(r)}e_{\varphi}$, $e_{\hat{z}} =\sqrt{1/D(r)}e_{z}\}$, 
for which $g_{_{\hat{\imath} \hat{\jmath}}} = \eta_{_{\hat{\imath}
\hat{\jmath}}}=diag(-1,1,1)$, we obtain
\begin{equation}
K_{\hat{\tau} \hat{\tau}}^{\pm} = \mp \frac{2A(a)B(a) \ddot{a}+A'(a)
+[A(a)B'(a)+A'(a)B(a)]\dot{a}^2}{2A(a)\sqrt{B(a)} \sqrt{1+B(a)\dot{a}^2}},
\label{e4}
\end{equation}
\begin{equation}
K_{\hat{\varphi} \hat{\varphi}}^{\pm} = \pm \frac{C'(a)\sqrt{1+B(a)
\dot{a}^2}}{2C(a)\sqrt{B(a)}}, 
\label{e5}
\end{equation}
and
\begin{equation}
K_{\hat{z} \hat{z}}^{\pm} = \pm \frac{D'(a)\sqrt{1+B(a)\dot{a}^2}}
{2D(a) \sqrt{B(a)}}, 
\label{e6}
\end{equation}
where the dot and the prime stand for $d / d \tau$ and $d/dr$ respectively. Therefore the surface energy density $\sigma=S_{\hat{\tau} \hat{\tau}}$ and the pressures $p_\varphi=S_{\hat{\varphi} \hat{\varphi}}$ and $p_z=S_{\hat{z} \hat{z}}$ for the matter on the shell are given 
by
\begin{equation}
\sigma = - \frac{\sqrt{1 + B(a) \dot{a}^2}}{8 \pi \sqrt{B(a)}} \left[
\frac{C'(a)}{C(a)} + \frac{D'(a)}{D(a)} \right],
\label{e8}
\end{equation}
\begin{equation}
p_{\varphi} =  \frac{1}{8 \pi \sqrt{B(a)} \sqrt{1 + B(a)\dot{a}^2}} 
\left\{ 2 B(a) \ddot{a} + B(a) \left[  \frac{A'(a)}{A(a)} +
\frac{ B'(a)}{B(a)} +\frac{D'(a)}{D(a)} \right] \dot{a}^2  + \frac{A'(a)}{A(a)} + \frac{D'(a)}{D(a)} \right\},
\label{e9}
\end{equation}
\begin{equation}
p_{z} =  \frac{1}{8 \pi \sqrt{B(a)} \sqrt{1 + B(a) \dot{a}^2}}
\left\{ 2 B(a) \ddot{a} + B(a) \left[ \frac{A'(a)}{A(a)} + \frac{B'(a)} {B(a)} +\frac{C'(a)}{C(a)} \right] \dot{a}^2 + \frac{A'(a)}{A(a)} + \frac{C'(a)}{C(a)} \right\} .
\label{e10}
\end{equation}
The definition of the wormhole throat usually applied to  compact configurations states that it is  a {\it minimal area surface}. In the cylindrical case, the area function ${\cal A}(r)=2\pi\sqrt{C(r)D(r)}$ should then increase at both sides of the throat, and this implies that $\lp CD \rp'(a)>0$. We can call it the {\it areal} flare-out condition. If $\lp CD \rp'(a)=C'(a)D(a)+C(a)D'(a)>0$, from Eq. (\ref{e8}) it follows that the surface energy density is negative, so the matter at the throat is \textit{exotic}. However,  Bronnikov and Lemos \cite{brle} recently introduced a new throat definition which only requires that  the circular radius function  ${\cal R}(r)=\sqrt{C(r)}$ has a minimum at the throat; thus, we can call this one  the {\it radial} flare-out condition. This definition implies $C'(a)>0$, imposing no condition on the sign of $(CD)'(a)$; in principle, depending on the metric considered, this could make possible a {\it positive} energy density.

In the case of the wormholes constructed from the spacetime given by Eq. (\ref{ds2}), by using the equations above, we find the energy density and the pressures
\begin{equation}
\sigma = -\frac{\sqrt{1+ \dot{a}^2}\sqrt{3|\Lambda |}  \left[2 \alpha (k) \cosh \left(\sqrt{3|\Lambda | }a\right)+ \zeta (k) \right]}{12 \pi \alpha (k) \sinh \left(\sqrt{3|\Lambda | }a\right)},
\end{equation}
\begin{equation}
p_\varphi =\frac{1}{4 \pi \sqrt{1+ \dot{a}^2}}\left\{\ddot{a}+\frac{\sqrt{3|\Lambda |} \left[2 \alpha (k) \cosh \left(  \sqrt{3|\Lambda |}a\right) - \epsilon(k) \right] (1+\dot{a}^2)}{3 \alpha (k) \sinh \left(\sqrt{3|\Lambda | }a\right)}\right\},
\end{equation}
\begin{equation}
p_z =\frac{1}{4 \pi \sqrt{1+ \dot{a}^2}}\left\{\ddot{a}+\frac{\sqrt{3|\Lambda |} \left[2 \alpha (k) \cosh \left( \sqrt{3|\Lambda |} a\right)- \gamma (k)\right] (1+\dot{a}^2)}{3 \alpha (k) \sinh \left(\sqrt{3|\Lambda | }a\right)}\right\}.
\end{equation}
Note that there is no dependence with the parameter $\xi$ of the metric. It is not difficult to see that the energy density at the shell is always negative for all the values of the parameter $k$ (in the range $0\le k \le 1/2$ adopted in this work) and for any throat radius $a$. This result is valid even if the radial flare-out condition is used. So the weak energy condition (which requires $\sigma \ge 0$ and $\sigma +p_j\ge 0$) cannot be fulfilled in our construction, and the matter at the throat is exotic.

For comparison, let us analyze the case of wormholes constructed from the Levi-Civita metric ($\Lambda=0$), for which we obtain that the energy density and the pressures are given by
\begin{equation}
\sigma (a)= -\frac{\sqrt{1+ \dot{a}^2}(1-2 k)^2}{4 \pi  a \left(4 k^2-2 k+1\right)},
\end{equation}
\begin{equation}
p_\varphi (a)=\frac{1}{4\pi \sqrt{1+ \dot{a}^2} }\left[ \ddot{a}+\frac{4 k^2 (1+\dot{a}^2)}{a (4 k^2-2  k+1)}\right],
\end{equation}
\begin{equation}
p_z (a)=\frac{1}{4 \pi\sqrt{1+ \dot{a}^2}} \left[ \ddot{a}+\frac{1+\dot{a}^2}{a(4k^2-2 k+1)}\right].
\end{equation}
The energy density at the shell is always negative for all the values of $k$ and $a$, even if the radial flare-out condition is adopted. Then, the weak energy condition is not satisfied, and the matter at the throat is also exotic.

\section{Stability}

The static values of the energy density and the pressures are obtained by putting $\dot{a}=0$ and $\ddot{a}=0$ in Eqs. (\ref{e8}), (\ref{e9}), and (\ref{e10}):
\begin{equation}
\sigma (a_0)= - \frac{1}{8 \pi \sqrt{B(a_0)}} \left[ \frac{C'(a_0)}{C(a_0)} + \frac{D'(a_0)}{D(a_0)} \right],
\label{e12}
\end{equation}
\begin{equation}
p_{\varphi}(a_0) =  \frac{1}{8 \pi \sqrt{B(a_0)}} \left[ \frac{A'(a_0)}{A(a_0)} +   \frac{D'(a_0)}{D(a_0)} \right],
\label{e13}
\end{equation}
\begin{equation}
p_{z} (a_0)=  \frac{1}{8 \pi \sqrt{B(a_0)}} 
\left[\frac{A'(a_0)}{A(a_0)} + \frac{C'(a_0)}{C(a_0)} \right] ,
\label{e14}
\end{equation}
where $a_0$ is the constant throat radius. Then, the static wormhole associated to the background geometry given by Eq. (\ref{ds2}) is supported by a shell at the throat with the surface energy density 
\begin{equation}
\sigma (a_0)=-\frac{\sqrt{3|\Lambda |}  \left[2 \alpha (k) \cosh \left(\sqrt{3|\Lambda | }a_0\right)+ \zeta (k) \right]}{12 \pi \alpha (k) \sinh \left(\sqrt{3|\Lambda | }a_0\right)},
\end{equation}
and the pressures
\begin{equation}
p_\varphi (a_0)=\frac{\sqrt{3|\Lambda |} \left[2 \alpha (k) \cosh \left(  \sqrt{3|\Lambda |}a_0\right) - \epsilon(k) \right] }{12 \pi \alpha (k) \sinh \left(\sqrt{3|\Lambda | }a_0\right)},
\end{equation}
\begin{equation}
p_z (a_0)=\frac{\sqrt{3|\Lambda |} \left[2 \alpha (k) \cosh \left( \sqrt{3|\Lambda |} a_0\right)- \gamma (k)\right] }{12 \pi \alpha (k) \sinh \left(\sqrt{3|\Lambda | }a_0\right)}.
\end{equation}

In the $\Lambda=0$ case, we find that
\begin{equation}
\sigma (a_0)=-\frac{(1-2 k)^2}{4 \pi  a_0 \left(4 k^2-2 k+1\right)},
\end{equation}
\begin{equation}
p_\varphi (a_0)=\frac{k^2}{\pi  a_0 (4 k^2-2  k+1)},
\end{equation}
\begin{equation}
p_z (a_0)=\frac{1}{4 \pi  a_0 (4k^2-2 k+1)}.
\end{equation}

In what follows, we adopt two possible formalisms for the stability analysis of the static configurations, previously presented in a general form in Refs. \cite{eisi10} and  \cite{mha}. We  consider small perturbations, so we assume that the geometry outside the throat remains static.

\subsection{Fixed equations of state}

A possible approach to the study of the mechanical stability relies on the fact that we are interested in small perturbations starting from a static  solution, so that the evolution of the shell matter can be considered as a succession of static states \cite{ei2,mc}. We summarize here our general analysis for radial perturbations introduced in Ref. \cite{ei1} and generalized in Ref.  \cite{eisi10}, in which we fix the form of the equations of state. By using Eq. (\ref{e12}), we see that Eqs. (\ref{e13}) and (\ref{e14}) can be recast in the form
\begin{equation}
p_{\varphi}(a_0)=-\frac{C(a_0)[A(a_0)D'(a_0)+A'(a_0)D(a_0)]}{A(a_0)[C(a_0)D'(a_0)+C'(a_0)D(a_0)]} \sigma (a_0),
\label{e15}
\end{equation}
\begin{equation}
p_{z}(a_0)=-\frac{D(a_0)[A(a_0)C'(a_0)+A'(a_0)C(a_0)]}{A(a_0)[C(a_0)D'(a_0)+C'(a_0)D(a_0)]} \sigma (a_0).
\label{e16}
\end{equation}
Then, the functions $A(a_0)$, $C(a_0)$, and $D(a_0)$ determine the equations of state 
$p_{\varphi}(\sigma)$ and $p_{z}(\sigma)$ of the exotic matter on the shell. Hence, we can assume that the equations of state for the dynamic case have the same form as in the static one, i.e. that they do not depend on the derivatives of $a(\tau)$; then $p_{\varphi}(\sigma)$ and $p_{z}(\sigma)$ are given by Eqs. (\ref{e15}) and (\ref{e16}) with $a$ instead of $a_0$. Therefore, as we have shown in our previous works, even in the most general case of a wormhole connecting two geometries of the form (\ref{e1}) a simple second order differential equation for $a(\tau )$ is obtained:
\begin{equation} 
2B(a)\ddot{a}+B'(a)\dot{a}^{2}=0.
\label{e17}
\end{equation}
Now, in our case for both $\Lambda < 0$ and $\Lambda = 0$, the metric  has $B(r)=1$, so that $\ddot a=0$, $\dot a(\tau)=\dot a(\tau_0)$ and then 
\begin{equation} 
a(\tau )=a(\tau_0 )+\dot{a}(\tau _{0})
(\tau - \tau _{0}).
\label{e20}
\end{equation}
So, when given in terms of the proper time on the shell, the radius of the wormhole throat can only undergo a monotonous evolution with constant speed. This result is a particular case of the general behavior found in our previous articles.   

\subsection{Linearized analysis}

Let us review the formalism introduced in Ref. \cite{mha}, which was previously developed for the spherically symmetric case in Refs. \cite{povi1,povi2,whsphe1,whsphe2}. From Eqs. (\ref{e8}), (\ref{e9}), and  (\ref{e10}), we can obtain the conservation equation
\begin{equation}
\frac{d(\mathcal{A}\sigma )}{d\tau} + p_{\varphi} \frac{\mathcal{A}}{\sqrt{C(a)}} \frac{d\sqrt{C(a)}}{d\tau} + p_z \frac{\mathcal{A}}{\sqrt{D(a)}} \frac{d\sqrt{D(a)}}{d\tau} = \dot{a} \mathcal{A} \mathcal{F}(a),
\label{ceq1}
\end{equation}
where $\mathcal{A}=2\pi \sqrt{C(a)D(a)}$ and 
\begin{equation*}
\mathcal{F}(a) = -\frac{\sigma }{2}\left\{ \frac{A'(a)}{A(a)}+\frac{B'(a)}{B(a)} +\left[ \frac{C'(a)^2}{C(a)^2}-\frac{2 C''(a)}{C(a)}+\frac{D'(a)^2}{D(a)^2}-\frac{2 D''(a)}{D(a)}\right] \left[ \frac{C'(a)}{C(a)}+\frac{D'(a)}{D(a)}\right]^{-1} \right\}.
\end{equation*}
In the left hand side of Eq. (\ref{ceq1}), the first term represents the internal energy change of the throat, and the second and the third terms represent the work done by the internal forces of the throat; while the right hand side corresponds to a flux. By using that $d/d\tau =\dot{a}d/da$, this conservation equation can be rewritten to give
\begin{equation}
(\mathcal{A}\sigma )' + \mathcal{A} p_{\varphi}  \frac{\left[\sqrt{C(a)}\right]'}{\sqrt{C(a)}} + \mathcal{A} p_z \frac{\left[\sqrt{D(a)}\right]'}{\sqrt{D(a)}} = \mathcal{A} \mathcal{F}(a).
\label{ceq2}
\end{equation}
The last equation can be formally integrated to obtain $\sigma = \sigma (a)$ if the pressures are known as functions of the energy density. From Eq. (\ref{e8}), we can write the equation of motion of the throat in the form
\begin{equation}
\dot{a}^2+V(a)=0,
\label{eqmo}
\end{equation}
with the potential
\begin{equation}
V(a)= B(a) ^{-1} - (8\pi)^2 \left[ \frac{C'(a)}{C(a)}+\frac{D'(a)}{D(a)}\right]^{-2} \sigma ^2 .
\label{pot}
\end{equation}
A Taylor expansion to second order of the potential $V(a)$ around the static solution gives
\begin{equation}
V(a)=V(a_{0})+V^{\prime }(a_{0})(a-a_{0})+\frac{V^{\prime \prime }(a_{0})}{2}
(a-a_{0})^{2}+O(a-a_{0})^{3}.
\label{taylor}
\end{equation}
It is easy to see that $V(a_0)=0$. Taking the derivative of the potential and using the conservation equation (\ref{ceq2}), we have that $V'(a_0)=0$. Using Eq. (\ref{ceq2}) again and introducing the parameters $\eta _1 = (dp_\varphi /d\sigma )(a_0) = p_\varphi '(a_0)/ \sigma'(a_0)$ and $\eta _2 = (dp_z/d\sigma )(a_0) = p_z' (a_0)/ \sigma'(a_0)$, the second derivative of the potential evaluated at $a_0$ reads
\begin{equation}
V''(a_0)= \frac{C'(a_0)D'(a_0)\chi (a_0) +\left[ C'(a_0)D(a_0) \eta_1 + C(a_0)D'(a_0)\eta_2 \right] \Omega (a_0)}{B(a_0)\left[C'(a_0)D(a_0)+C(a_0)D'(a_0)\right]},
\label{potd2}
\end{equation}
where
\begin{eqnarray}
\chi (a_0) &=&  \frac{A'(a_0)}{A(a_0)} \left[ \frac{C(a_0)}{C'(a_0)}+\frac{D(a_0)}{D'(a_0)}\right]  \left[ -\frac{A'(a_0)}{A(a_0)}+\frac{A''(a_0)}{A'(a_0)}-\frac{B'(a_0)}{2  B(a_0)}\right] \nonumber \\
&& -\frac{B'(a_0)}{B(a_0)}-\frac{C'(a_0)}{C(a_0)}+\frac{C''(a_0)}{C'(a_0)}-\frac{D'(a_0)}{D(a_0)}+  \frac{D''(a_0)}{D'(a_0)}, \nonumber
\end{eqnarray}
and
\begin{equation*}
\Omega (a_0)=\frac{C'(a_0)}{C(a_0)} \left[ -\frac{B'(a_0)}{2 B(a_0)}-\frac{C'(a_0)}{C(a_0)}+\frac{C''(a_0)}{C'(a_0)}\right] + \frac{D'(a_0)}{D(a_0)} \left[ -\frac{B'(a_0)}{2 B(a_0)}-\frac{D'(a_0)}{D(a_0)}+\frac{D''(a_0)}{D'(a_0)}\right].
\end{equation*}
A static solution is stable under radial perturbations if $V''(a_0)>0$. So the stability analysis reduces to the study of the sign of the second derivative of the potential.
  
\begin{figure}[t!]
\centering
\includegraphics[width=12cm]{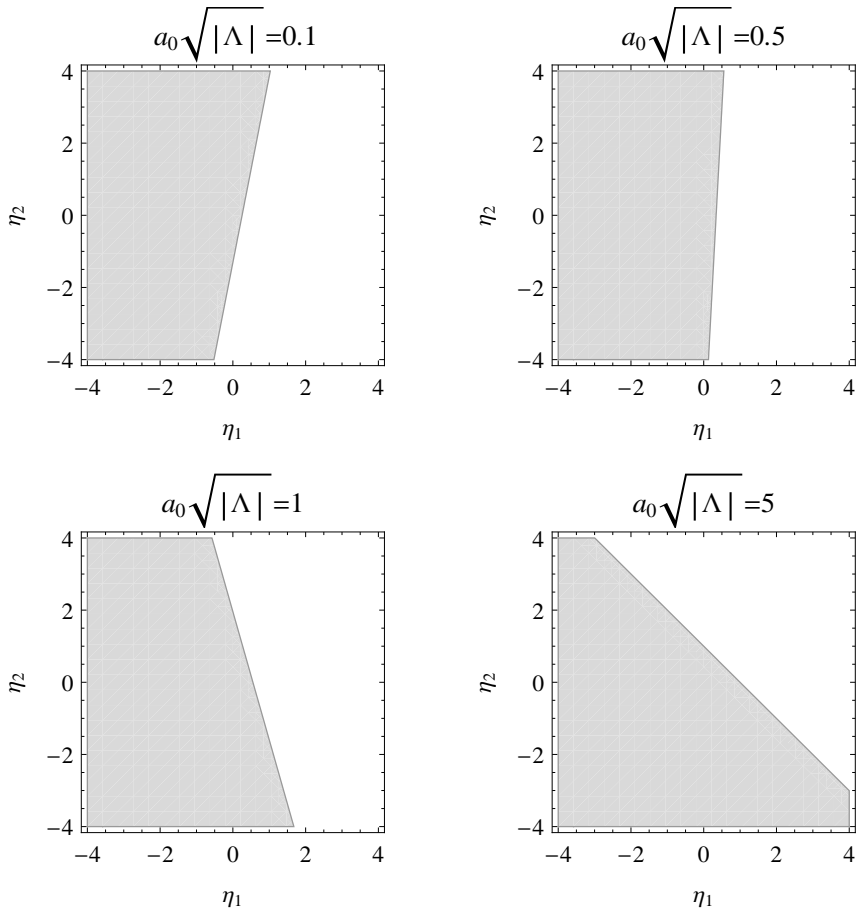}
\caption{Thin-shell wormholes constructed from the Linet--Tian metric with $\Lambda<0$: linearized stability regions (in gray) in the plane $(\eta _1, \eta _2)$,  for $k=0.1$ and several values of $a_0 \sqrt{|\Lambda|}$.}\label{f1}
\end{figure}

\begin{figure}[t!]
\centering
\includegraphics[width=12cm]{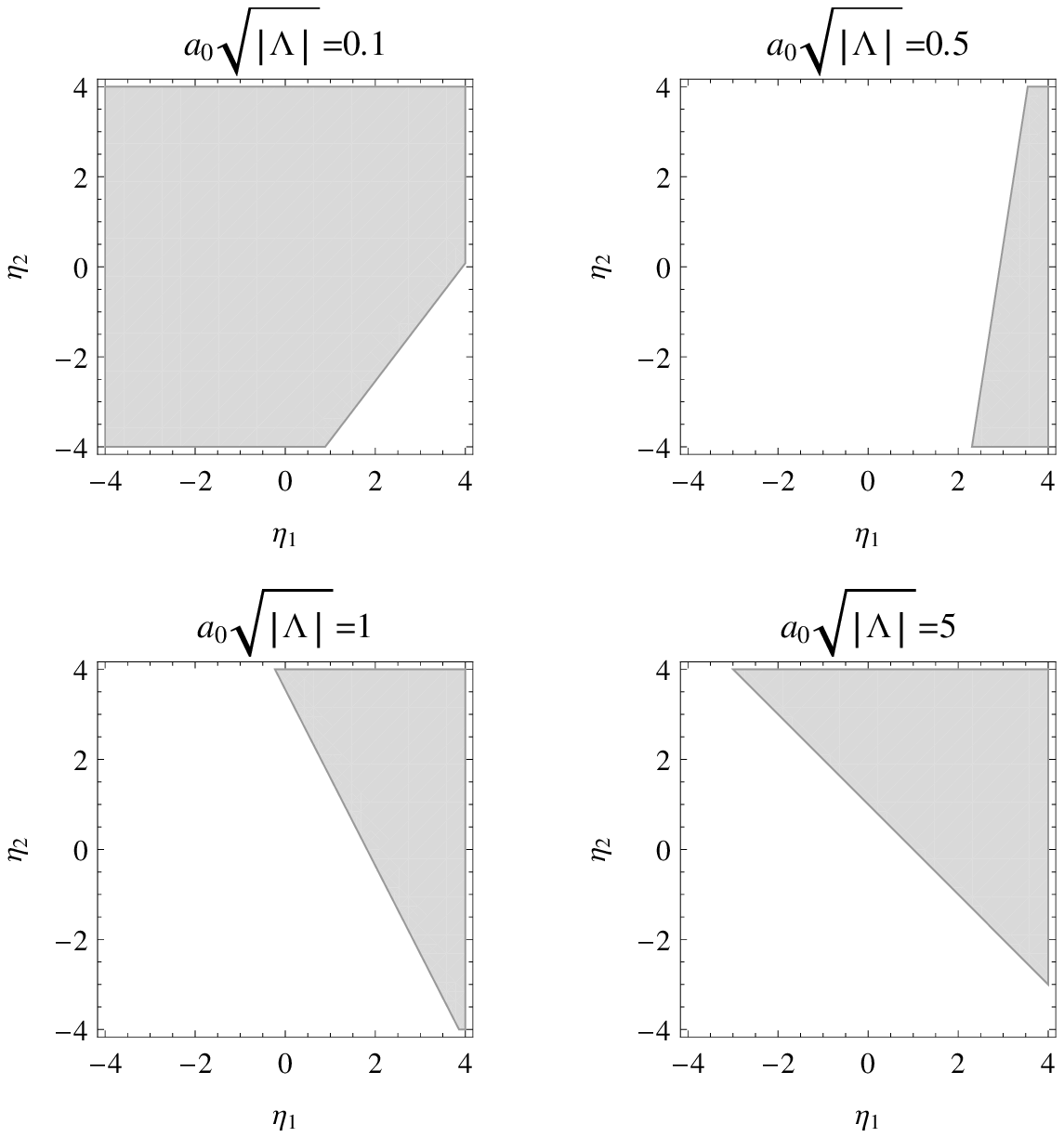}
\caption{Thin-shell wormholes constructed from the Linet--Tian metric with $\Lambda<0$: linearized stability regions (in gray) in the plane $(\eta _1, \eta _2)$, for $k=0.4$ and several values of $a_0 \sqrt{|\Lambda|}$.}\label{f2}
\end{figure}

\begin{figure}[t!]
\centering
\includegraphics[width=12cm]{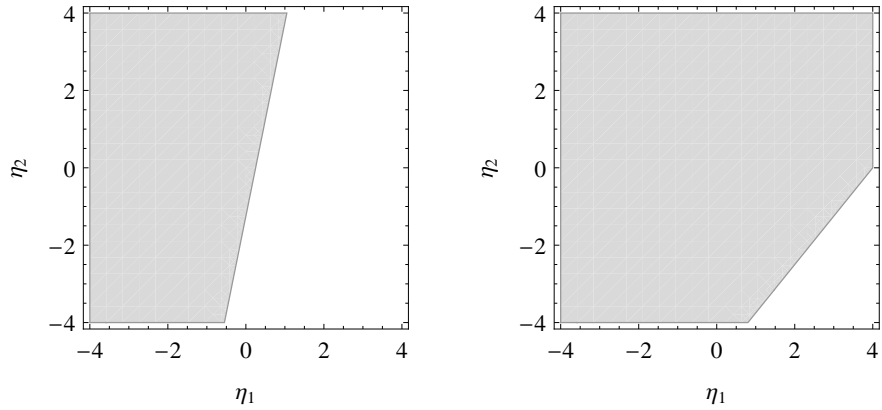}
\caption{Thin-shell wormholes constructed from the Levi-Civita metric ($\Lambda=0$): linearized stability regions (in gray) in the plane $(\eta _1, \eta _2)$, for $k=0.1$ (left) and for $k=0.4$ (right). The results do not depend on $a_0$ (see text).}\label{f3}
\end{figure}

For the static wormhole constructed from the metric given by Eq. (\ref{ds2}), we obtain that 
\begin{eqnarray}
V''(a_0) &=&\frac{2 \Lambda  \left[\zeta (k) \cosh \left( a_0 \sqrt{3|\Lambda | }\right)+2 \alpha (k)\right] } {\alpha (k) \left[ 2 \alpha (k) \cosh \left( a_0 \sqrt{3|\Lambda | }\right)+\zeta (k)\right] \sinh^2\left( a_0 \sqrt{3|\Lambda | }\right) }
 \nonumber \\
&& \times \left[ (\eta _1+\eta _2-1) \alpha (k) \cosh \left( a_0 \sqrt{3|\Lambda | }\right)+ \eta _1 \epsilon (k) +\eta _2 \gamma(k)+\zeta (k)\right] .
\end{eqnarray}
The functions in the denominator $\alpha (k)$ and $\sinh^2( a_0 \sqrt{3|\Lambda | })$ are always positive, and the function $ 2 \alpha (k) \cosh ( a_0 \sqrt{3|\Lambda | })+\zeta (k)$ is also positive for all $k$ within the range $0< k < 1/2$, so using that $\Lambda<0$ we can simplify the condition for $V''(a_0)>0$ in the form 
\begin{equation}
\left[\zeta (k) \cosh \left( a_0 \sqrt{3|\Lambda | }\right)+2 \alpha (k)\right] \left[ (\eta _1+\eta _2-1) \alpha (k) \cosh \left( a_0 \sqrt{3|\Lambda | }\right)+ \eta _1 \epsilon (k) +\eta _2 \gamma(k)+\zeta (k)\right] < 0.
\end{equation}
We see that this inequality is rather complicated to be interpreted analytically. So we show the results graphically in Figs. \ref{f1} and \ref{f2}, in the plane $(\eta _1, \eta _2)$, for some relevant values of the other parameters. For given values of $\Lambda$, $k$, and $a_0$, in each plot the stable region (in gray) is limited by the straight line
\begin{equation}
\left[ \alpha (k) \cosh \left( a_0 \sqrt{3|\Lambda | }\right)+ \epsilon (k) \right] \eta _1  + \left[ \alpha (k) \cosh \left( a_0 \sqrt{3|\Lambda | }\right)+ \gamma(k)  \right] \eta _2 = \alpha (k) \cosh \left( a_0 \sqrt{3|\Lambda | }\right)-\zeta (k),
\end{equation}
which has a positive slope for small values of $a_0 \sqrt{|\Lambda|}$; this slope becomes negative as $a_0 \sqrt{|\Lambda|}$ increases. If the stable region lies above or below this line depends also on the sign of the expression $\zeta (k) \cosh ( a_0 \sqrt{3|\Lambda | })+2 \alpha (k)$, which is positive for small $k$ and becomes negative as $k$ grows (the value where it changes sign is determined by $a_0 |\Lambda |$). The plots corresponding to small values of $k$, exemplified by Fig. \ref{f1} for $k=0.1$, show that the stable region is above the straight line for small values of $a_0 \sqrt{|\Lambda|}$ and is below this line as $a_0 \sqrt{|\Lambda|}$  grows. 
When $k$ is large, as displayed in Fig. \ref{f2} for $k=0.4$, the stable region is above the limiting line for small $a_0 \sqrt{|\Lambda|}$, below this line when $a_0 \sqrt{|\Lambda|}$ grows, and then above again. 
From the figures we see that, for appropriate values of $k$ and $a_0 \sqrt{|\Lambda|}$, stable configurations can be found with both $\eta_1$ and $\eta_2$ belonging to the interval $[0,1]$, which is an useful feature if one wants to interpret these parameters as the squared velocity of sound in the corresponding directions. But for the exotic matter at the throat, this interpretation is not mandatory \cite{povi2}. 

In the case of  wormholes constructed from the Levi-Civita geometry, we find that the second derivative of the potential takes the form
\begin{equation}
V''(a_0)=\frac{-2 \eta _1 - 8 k^2 \eta _2 + 4 k (\eta _1+\eta _2+1)}{a_0^2 \left(4 k^2-2 k+1\right)},
\end{equation}
where $0<k<1/2$, as stated previously. The denominator is always positive, so the stability condition $V''(a_0)>0$ reduces to 
\begin{equation}
 -(1-2k) \eta _1 + 2k (1-2k) \eta _2 + 2k >0,
\end{equation}
which is independent of the throat radius $a_0$. We show the results  in the plane $(\eta _1, \eta _2)$ for different values of the other parameters in Fig. \ref{f3}. We see that, for a given value of $k$ (two values: $0.1$ and $0.4$ are shown in the plots), there is a stable region above the line 
\begin{equation}
-(1-2k) \eta _1 + 2k (1-2k) \eta _2 = -2k .
\end{equation}
It is easy to verify that the slope of this line is positive and a decreasing function of $k$. Configurations with both $\eta_1$ and $\eta_2$ within the interval $[0,1]$ are possible for all values of $k$ in the range adopted.

\section{Discussion}\label{discu}

We have constructed cylindrical thin-shell wormholes with anti-de Sitter asymptotics, starting from the Linet--Tian metric with $\Lambda <0$. We have characterized the matter at the throat, finding that it cannot satisfy the energy conditions for any of the two usual flare-out definitions; i.e. it is always exotic. We have also studied the case with $\Lambda = 0$,  corresponding to wormholes constructed from the Levi-Civita spacetime, which are also always supported by exotic matter at the throat. We have  performed two different stability analyses under perturbations preserving the symmetry. If we fix the form of the equations of state by assuming that those corresponding to a dynamic throat maintain the form of the static case, the configurations are unstable, as it was previously shown in the general case \cite{eisi10}. Instead, in the linearized stability analysis, stable configurations are possible for suitable values of the parameters. 
When $\Lambda=0$ the stability does not depend on $a_0$, while if $\Lambda<0$ it depends on the product $a_0 \sqrt{|\Lambda|}$. For $\Lambda=0$ and $\Lambda<0$, stable configurations with both $0 \le \eta _1 \le 1$ and $0 \le \eta _2 \le 1$ are possible, so these parameters can be interpreted as the squared velocity of sound in the corresponding directions. In the $\Lambda < 0$ case, this desirable feature is present for any value of $a_0 \sqrt{|\Lambda|}$  when $k$ is small and only for low values of $a_0 \sqrt{|\Lambda|}$ if $k$ is large. 

\section*{Acknowledgments}

This work has been supported by Universidad de Buenos Aires and CONICET.

\end{document}